\documentclass[preprint,12pt]{elsarticle}

\def\lbldef#1#2{\expandafter\gdef\csname #1\endcsname {#2}}

\def\href#1#2{#2}

\usepackage{color}
\usepackage{graphicx}
\usepackage{amssymb}
\journal{Physics of the Dark Universe}

\begin{document}
	
	\begin{frontmatter}

		\title{Structure formation and the
				matter power-spectrum in the $R_{\rm h}=ct$ universe}

		\author[1]{Manoj K. Yennapureddy\footnote{E-mail: manojy@email.arizona.edu}} 
		\author[2]{Fulvio Melia\footnote{John Woodruff Simpson
				Fellow. E-mail: fmelia@email.arizona.edu}} 
		
		\address[1]{Department of Physics, The University of Arizona, AZ 85721, USA}
		\address[2]{Department of Physics, The Applied Math Program, and Department of Astronomy,
			The University of Arizona, AZ 85721, USA}
		
		\begin{abstract}
			Inflation drives quantum fluctuations beyond the Hubble horizon, freezing them out before the small-scale modes re-enter during the radiation dominated epoch, and subsequently decay, while large-scale modes re-enter later during the matter dominated epoch and grow. This distinction shapes the matter power spectrum and provides observational evidence in support of the standard model. In this paper, we demonstrate that another mechanism, based on the fluctuation growth in the $R_{\rm h}=ct$ universe, itself an FLRW cosmology with the added constraint of zero active mass (i.e., $\rho+3p=0$), also accounts very well for the observed matter power spectrum, so this feature is not unique to $\Lambda$CDM. In $R_{\rm h}=ct$, the shape of the matter power spectrum is set by the interplay between the more rapid decay of the gravitational potential for the smaller mode wavelengths and the longer dynamical timescale for the larger wavelengths. This combination produces a characteristic peak that grows in both amplitude and mode number as a function of time. Today, that peak lies at $k\approx 0.02$ Mpc$^{-1}$, in agreement with the Ly-$\alpha$ and {\it Planck} data. But there is no need of an inflationary expansion, and a complicated epoch dependence as one finds in $\Lambda$CDM.
		\end{abstract}
		
	\end{frontmatter}
	
\section{Introduction}
Our current view of cosmic evolution holds that large-scale structure originated from primordial quantum fluctuations, believed to have been seeded during inflation in the early Universe \cite{Guth1981,Mukahnov1981,Guth1982,Hawking1982,Starobinsky1982,Mukhanov2005}. According to this theory, quantum fluctuations classicalized and grew due to self-gravity to form the inhomogeneous Universe we see today. The existence of these inhomogeneities in the CMB and matter power spectrum \cite{Bennett2003,Spergel2003,Tegmark2004} has become quite evident with the advent of high precision measurements. Moreover, the observation of galaxy rotation curves \cite{Zwicky1937,Sofue1999,Combes2002} and weak lensing measurements \cite{Hammer1997,Kneib1996,Gioia1998,Tyson1998,Katherine2017} have indicated a necessity for the existence of dark matter.  Hence primordial fluctuations must include dark and baryonic matter, and radiation. Dark matter is assumed to be collisionless and non-interacting, so its fluctuations grew due to self-gravity only, without the suppression from radiation pressure. Once dark matter perturbations grew beyond a critical limit, they formed bound objects \cite{PressSchechter1974}. In contrast, baryons could not form bound objects in the early Universe because their growth was suppressed by the radiation to which they were coupled. As the baryons gradually decoupled from the radiation, however, they accreted into the potential wells created by dark matter to form the large-scale structure we see in the Universe today. This scenario---in which dark matter participated in the formation of structure in the early Universe---is indispensable to $\Lambda$CDM. Without the dark matter, structure formation would have been delayed by the baryon-radiation coupling, creating an inconsistency with the observations.

But the high precision data coming down the pipeline over the past decade have created a conflict with the timeline of structure formation in $\Lambda$CDM, even with the contribution due to dark matter. In particular, the discovery of SDSS010013.02+280225.8, an ultraluminous quasar at redshift $z=6.3$ \cite{Wu2015}, has exacerbated the problem of supermassive black-hole growth and evolution in the early Universe \cite{MeliaMcclintock2015a}. Almost all of the 122 previously discovered quasars at redshifts $z\approx6$ \cite{Fan2003,Jiang2007,Williott2007,Jiang2008,Williott2010,Mortlock2011,Banados2014,Banados2016,Jiang2016,Matsuoka2016,Mazzucchelli2017,Tang2017,Wang2017,Chehade2018,Matsuoka2018,Pons2019,Shen2019} have a black hole with mass $\sim10^9\;M_{\odot}$, challenging the standard model's predicted timeline, which would have afforded them less than 900 Myr to  grow after the big bang, but likely even less than $\sim$ 500 Myr since the onset of Population II star formation. 

In the context of $\Lambda$CDM, the formation of Pop III stars might have occurred by redshift  $z\approx 20$ \cite{Haiman1996,Tegmark1997,Abel2002,Bromm2002}, corresponding to a time of 200 Myrs. Then the subsequent transition to Pop II stars would have taken at least 100 Myrs because the gas expelled by the first generation stars had to cool and recollapse \cite{Yoshida2004, Johnson2007}. These Pop II stars would have acted as seeds for black holes. If the early black holes grew according to the Eddington rate, it would have taken at least 930 Myrs for them to reach their observed mass. The observed quasar at redshift $z = 6.3$, however, would have been 880 Myrs old in $\Lambda$CDM, which is not consistent with how these objects grew. 

Two possibilities have been proposed to reconcile this problem: 1) the black holes might have grown with highly anomalous accretion rates, but this phenomenon is not observed anywhere in the Universe; 2) Pop III stars might have formed earlier than expected, thus partially reconciling the problem. But this too doesn't appear to be reasonable according to various simulations \cite{Barkana2001,Miralda2003,Bromm2004,Ciardi2005,Glover2005,Grief2007,Wise2008,Bromm2009,Salvaterra2011,Grief2012,Jaacks2012,Yoshida2012}. In addition, the structure formation timeline in $\Lambda$CDM predicts that at redshifts $z = 4 - 8$ most massive galaxies should have been transitioning from an initial halo assembly to baryonic evolution. But the recent Cosmic Assembly Near-infrared Deep Extragalactic Survey (CANDELS)\cite{Gorgin2011,Koekemoer2011} and Spitzer Large Area Survey with Hyper-Suprime-Cam (SPLASH) surveys have found that these massive halos formed much earlier than predicted by $\Lambda$CDM, giving to rise to what some have called `The Impossibly Early Galaxy Problem' \cite{Steinhardt2016,YennapureddyMelia2018a,Yennapureddy2019}. Attempts to reconcile this problem have indicated that a $\sim 0.8$ dex change is required in the halo to stellar mass ratio over this redshift range, though such a drastic change could come about only with a complete absence of dark matter at redshift 8 or with essentially $100\%$ of the baryons condensing into stars at higher redshifts \cite{Steinhardt2016}. Both of these scenarios constitute implausible physics, such as the need to convert all of the baryons into stars instantly upon halo virialization (see ref. \cite{Steinhardt2016} and references cited therein). Other attempted remedies, such as an evolution of the halo mass to light ratio, could reconcile the problem, but could only happen if the initial mass function were top-heavy. Studies have shown that the ratio of halo mass to light should remain the same, at least within the redshift range ($4\leq z \leq 8$).

In this paper, we present the formation of structure in $R_{\rm h}=ct$ \cite{Melia2003,Melia2007,MeliaShevchuk2012,Melia2013a,Melia2016a,Melia2016b}, an alternative Friedmann-Lema\^itre-Robertson-Walker (FLRW) cosmology with zero active mass \cite{Melia2016a,Melia2016b}. The linear expansion in this model afforded sufficient time for supermassive objects to form, thus solving the `too early' appearance of massive quasars \cite{MeliaMcclintock2015a}. Even more importantly, we recently showed that the growth rate in this model accounts very well for the (otherwise too) early appearance of massive halos and galaxies \cite{YennapureddyMelia2018a,Yennapureddy2019}. 

In \S~2, we summarize the basis for the two cosmological models we consider here (i.e., $\Lambda$CDM and $R_{\rm h}=ct$). Then, in \S\S~3--5, we shall derive the necessary mathematical formalism for the formation of structure in the $R_{\rm h}=ct$ universe, and describe the evolution of the matter power spectrum in \S\S~6--8. Our conclusions will be presented in \S~9.

\section{The Cosmological Models}
\subsection{Inflationary $\Lambda$CDM}
The standard model contains dark energy, dark matter, radiation and baryons as the primary constituents. The existence of dark energy is inferred from the distance-redshift relation of Type Ia SNe \cite{Perlmutter1998,Riess1998,Schmidt1998}, while the presence of dark matter is inferred from the galaxy  rotation \cite{Zwicky1937,Sofue1999,Combes2002} and weak lensing \cite{Hammer1997,Kneib1996,Gioia1998,Tyson1998} measurements, as noted earlier. Its current energy budget is dominated by dark energy ($\sim 70\%$; see ref. \cite{PlanckCollaboration2016}), possibly in the form of a cosmological constant with an equation-of-state $w_{\rm de}\equiv p_{\rm de}/\rho_{\rm de}=-1$. The dark-matter component contributes an additional $\sim 27\%$ of the energy budget, and is primarily responsible for the formation and growth of large-scale structures. The rest of the energy density ($\sim 3\%$) is in the form of baryons, with a negligible contribution from radiation.

The early $\Lambda$CDM universe was radiation dominated, producing a phase of decelerated expansion, leading to the well-known temperature horizon problem. The standard model is therefore strongly dependent on an early inflationary expansion, lasting from $10^{-35}$ to $10^{-32}$ s, in order to mitigate such internal inconsistencies \cite{Guth1981}. In more recent times, however, the principal benefit of inflation has been viewed as the mechanism it fosters for the creation and growth of quantum fluctuations \cite{Mukahnov1981,Guth1982,Hawking1982,Starobinsky1982} that might haved subsequently grown
into large-scale inhomogeneities. As stated earlier, the currently held view is that the classicalised 
version of these quantum fluctuations produced large-scale structure. Although, the actual mechanism of 
classicalisation has remained elusive \cite{Penrose2004, Mukhanov2005,Perez2006,Weinberg2008,LythLiddle2009,Bengochea2015}, it is thought that primordial 
modes were driven beyond the Hubble horizon by inflation, frozen and turned into classical modes. They
re-entered the horizon once the Hubble radius grew sufficiently. This exit and entry of modes across
the horizon is quite critical, playing a vital role in producing the matter power spectrum that we
shall describe in more detail in \S~7.

\subsection{The $R_{\rm h}=ct$ Universe}
Our primary focus in this paper is to demonstrate how the $R_{\rm h}=ct$ universe also provides a
mechanism for producing the observed matter power spectrum. This is an FLRW cosmology constrained
by the zero active mass condition, which results in an expansion with an apparent (or gravitational)
radius always equal to $ct$ \cite{Melia2003,Melia2007,Melia2012,Melia2013a,Melia2016a,Melia2016b,MeliaAbdelqader2009}. It too assumes the presence of dark energy, radiation and baryonic and dark matter. 
The primary difference between $\Lambda$CDM and $R_{\rm h}=ct$ is that the latter is constrained by 
the equation-of-state $\rho+3p=0$ \cite{Melia2016a,Melia2016b}, where $p$ is the total pressure and 
$\rho$ is the total energy density. Numerous comparative tests examining which of the two 
models, $\Lambda$CDM or $R_{\rm h}=ct$, is favoured by the data have been carried out over the 
past decade. These include high-$z$ quasars \cite{MeliaMcclintock2015b,Melia2013a,KauffmannHaehnelt2000,WyitheLoeb2003,Melia2014,Melia2018}, gamma ray bursts 
\cite{Dai2004,Ghirlanda2004,Wei2013}, cosmic chronometers \cite{MeliaMcclintock2015b,Melia2018,JimenezLoeb2002,Simon2005,MeliaMaier2013}, Type Ia SNe 
\cite{Melia2012,Perlmutter1998,Riess1998,Schmidt1998,Wei2015b}, Type Ic superluminous SNe \cite{Inserra2014,Wei2015a}, and the 
age measurements of passively evolving galaxies \cite{Alcaniz1999,LimaAlcaniz2000,Wei2015c}, and strong gravitational lensing \cite{YennapureddyMelia2018b,Melia2015,LeafMelia2018b}.

In all such one-on-one comparisons completed thus far, model selection tools show 
that the data favour $R_{\rm h}=ct$ over $\Lambda$CDM \cite{Melia2013b,Melia2014,MeliaMaier2013,MeliaMcclintock2015a,MeliaMcclintock2015b,Wei2013,Wei2015a,Wei2015b,Wei2015c}. What is particularly attractive
about this alternative FLRW cosmology is that it did not have an early decelerated or accelerated phase, 
so it has no horizon problem \cite{Melia2013b}. An inflationary phase is not needed in the $R_{\rm h}=ct$ 
universe, removing the need to find a self-consistent theory of inflation, which has thus far eluded
us for over three decades. The early $R_{\rm h}=ct$ universe was dominated by dark energy and radiation,
in the fractions $\sim 80\%$ and $\sim 20\%$ of the total energy density, respectively \cite{MeliaFatuzo2016}. The late $R_{\rm h}=ct$ universe is dominated by dark energy and matter, comprising
2/3 and 1/3 of the energy density, respectively.  In addition, since the ratio of horizon size to
proper mode size always remains constant in the $R_{\rm h}=ct$ universe, modes never cross the horizon.  
Thus, the mechanism for generating the matter power spectrum in $R_{\rm h}=ct$ is completely different 
than that in $\Lambda$CDM.

\section{Perturbed Einstein Equations in FLRW}
We begin by deriving the perturbed Einstein equations for an arbitrary FLRW metric and then use them in 
the $R_{\rm h}=ct$ universe to obtain the growth rate and power spectrum at various epochs. All the modes 
remain sub-horizon in $R_{\rm h}=ct$, but one has to use relativistic perturbation theory rather than 
Newtonian theory because the cosmic fluid in $R_{\rm h}=ct$ is always dominated by dark energy. We proceed 
by perturbing the general FLRW metric \cite{Weinberg1972,LandauLifshitz1975,Peebles1980,Peebles1993,KolbTurner1990, Padmanabhan1993,ColesLucchin1995,Peacock1999,LiddleLyth2000,Tsagas2008}, which we write in the form 
\begin{equation}
ds^2=a^2(\eta)\bigg[-(1-\bar{h}_{00})\,d\eta^2+2\bar{h}_{0\alpha}\,d\eta\,dx^\alpha+
(\delta_{\alpha \beta}+2\bar{h}_{\alpha\beta})\,dx^\alpha\,dx^\beta\bigg]\;,
\end{equation} 
where $a(\eta)$ is the scale factor in terms of the conformal time $\eta$. The components of the tensor 
$\bar{h}$ are the perturbations to the homogeneous FLRW metric. The indices $\alpha$ and $\beta$ run from 1 
to 3. Using the conventional scalar-vector-tensor decomposition, we write $\bar{h}_{00}$, $\bar{h}_{0\alpha}$ 
and $\bar{h}_{\alpha\beta}$ as follows
\begin{equation}
\bar{h}_{00}=2\Phi\;,
\end{equation}
\begin{equation}
\bar{h}_{0\alpha}=w_\alpha=w_\alpha^\perp+\partial_\alpha\phi^\parallel\;,
\end{equation}
and
\begin{equation}
\bar{h}_{\alpha \beta}=\bigg[ -\psi \delta_{\alpha \beta}+\bigg(\nabla_{\alpha}U_{\beta}^{\perp}+
\nabla_{\beta}U_{\alpha}^{\perp}\bigg)+
\bigg(\nabla_{\alpha}\nabla_{\beta}-\frac{1}{3}\delta_{\alpha \beta} 
\nabla^2\bigg)\Phi_1+h_{\alpha \beta}^{\perp \perp}\bigg]\;,
\end{equation}
with $\psi={h^\alpha}_\alpha$ and $h_\alpha^{\perp\perp {\alpha}}=0$. All these metric perturbations are 
highly coordinate dependent, so one cannot automatically attribute any physical meaning to the perturbed 
components of the metric. 

For infinitesimal coordinate transformations, such as $\bar{x}^i=x^i+\xi^i$, the decomposed perturbed 
metric components transform as follows
\begin{equation}
\bar{\Phi}=\Phi-\frac{1}{a}\frac{da}{d\eta}\xi^0\;,
\end{equation}
\begin{equation}
\bar{\phi}^\parallel=\phi^\parallel+\xi^0-\frac{d\xi}{d\eta}\;,
\end{equation}
\begin{equation}
\bar{w}^\perp_\alpha=w^\perp_\alpha-\frac{d\xi^{\perp}_\alpha}{d\eta}\;,
\end{equation}
\begin{equation}
\bar{\psi}=\psi+\frac{1}{a}\frac{da}{d\eta}\xi^0+\frac{1}{3}\nabla^2\xi\;,
\end{equation}
\begin{equation}
\bar{\Phi}_1=\Phi_1-\xi\;,
\end{equation}
\begin{equation}
\bar{U}_\alpha^\perp=U_\alpha^\perp-\frac{1}{2}\xi_\alpha^\perp\;,
\end{equation}
and
\begin{equation}
\bar{h}^{\perp\perp}_{\alpha\beta}=h^{\perp\perp}_{\alpha\beta}\;.
\end{equation}
To use these perturbed components, one may either choose a particular observer (i.e., a particular set of
coordinates), resort to a particular gauge, or work with Bardeen's four gauge independent variables 
\cite{Bardeen1980}. It is evident from Equations~(5)--(11) that there are seven independent metric perturbations, 
implying seven degrees of freedom, but there are only four gauge-invariant variables. So when choosing a 
particular gauge, one must fix 3 degrees of freedom. In this work, we follow the first approach by choosing 
the Newtonian gauge and fix three independent metric perturbations as follows (with
$\phi^{\parallel}=\Phi_1=U_{\alpha}^\perp=0$):
\begin{equation}
\Phi_A=\Phi+\frac{1}{a}\frac{\partial[a(\phi^\parallel-\frac{d\Phi_1}{d\eta})]}{\partial \eta}\;,
\end{equation}
\begin{equation}
\Phi_H=\frac{1}{a}\frac{da}{d\eta}(\phi^\parallel-\frac{d\Phi_1}{d\eta})-\psi-\frac{1}{3}
\partial^\alpha\partial_\alpha\Phi_1\;,
\end{equation}
\begin{equation}
\psi_\alpha=w_\alpha^\perp-2\frac{dU_\alpha^{\perp}}{d\eta}\;,
\end{equation}
and
\begin{equation}
h_{\alpha \beta}^\perp=h_{\alpha \beta}^\perp\;.
\end{equation}
After setting three metric perturbations to zero (in the Newtonian gauge), one ends up with 
\begin{equation}
\Phi_A=\Phi\;,
\end{equation}
\begin{equation}
\Phi_H=-\psi\;,
\end{equation}
\begin{equation}
\psi_\alpha=w_\alpha^\perp\;,
\end{equation}
and
\begin{equation}
\bar{h}_{\alpha \beta}^\perp=h_{\alpha \beta}^\perp\;.
\end{equation}
It is quite evident from the above equations that, in the Newtonian gauge, the independent metric 
perturbations are in fact the Bardeen variables. This is the primary reason for choosing this gauge. 
Then the perturbed metric is given as 
\begin{eqnarray}
\qquad ds^2&=&a^2(\eta)\left[-(1+2\Phi)\,d\eta^2+2w_\alpha^\perp\,d\eta\,dx^\alpha+\right.\nonumber\\
&\null&\qquad\qquad\left.\left\{(1-2\psi)\delta_{\alpha \beta}+ 
2h_{\alpha\beta}^{\perp \perp}\right\}\,dx^\alpha\,dx^\beta\right]\;.
\end{eqnarray}

It is well known that the vector perturbations die away as the Universe expands, and since our
primary motivation in this paper is to compute the power spectrum and evolution of the perturbed 
dark-matter density field, we shall ignore both the vector and tensor perturbations for this work. 
The metric with only the scalar perturbations is then given as 
\begin{equation}
ds^2=a^2(\eta)\bigg[-(1+2\Phi)\,d\eta^2+(1-2\psi)\delta_{\alpha\beta}\,dx^\alpha\,dx^\beta\bigg]\;,
\end{equation} 
which may be used to compute the Ricci tensor with the following components:
\begin{equation}
R_{00}=\nabla^2\Phi+3\frac{d^2\psi}{d\eta^2}+3{\mathcal{H}}\bigg[\frac{d\Phi}{d\eta}+
\frac{d\psi}{d\eta}\bigg]-3\frac{d{\mathcal{H}}}{d\eta}\;,
\end{equation}
\begin{equation}
R_{0\alpha}=2\partial_\alpha \frac{d\psi}{d\eta}+2{\mathcal{H}}\partial_\alpha\Phi\;,
\end{equation}
\begin{eqnarray}
R_{\alpha\beta}&=&\left[\frac{d{\mathcal{H}}}{d\eta}+2{\mathcal{H}}^2-\frac{d^2\psi}{d\eta^2}+
\nabla^2\psi-2\left(\frac{d{\mathcal{H}}}{d\eta}
+2{\mathcal{H}}^2\right)(\psi+\Phi)-\right.\nonumber\\
&\null&\left.{\mathcal{H}}\frac{d\Phi}{d\eta}-
5{\mathcal{H}}\frac{d\psi}{d\eta}\right]\delta_{\alpha\beta}
+\partial_\alpha\partial_\beta(\psi+\Phi)\;,
\end{eqnarray}
the Ricci scalar,
\begin{eqnarray}
a^2{\mathcal{R}}&=&6\left(\frac{d{\mathcal{H}}}{d\eta}+{\mathcal{H}}^2\right)-2\nabla^2\Phi+4\nabla^2\psi-
12\left(\frac{{\mathcal{H}}}{d\eta}
+{\mathcal{H}}^2\right)\Phi-\nonumber\\
&\null&6\frac{d^2\psi}{d\eta^2}-6{\mathcal{H}}\left(\frac{d\Phi}{d\eta}+3\frac{d^2\psi}{d\eta^2}\right)\;,\;\;
\end{eqnarray}
and also the Einstein tensor,
\begin{equation}
G_{00}=3{\mathcal{H}}^2+2\nabla^2\psi-6{\mathcal{H}}\frac{d\psi}{d\eta}\;,
\end{equation}
\begin{equation}
G_{0\alpha}=2\partial_\alpha\frac{d\psi}{d\eta}+2{\mathcal{H}}\partial_\alpha\Phi\;,
\end{equation}
\begin{eqnarray}
G_{\alpha\beta}&\hskip-0.2in=\hskip-0.2in&-\left(2\frac{d{\mathcal{H}}}{d\eta}+{\mathcal{H}}^2\right)
\delta_{\alpha\beta}+\left[\nabla^2(\Phi-\psi)+2\frac{d^2\psi}{d\eta^2}
+2\left(2\frac{d{\mathcal{H}}}{d\eta}+\right.\right.\nonumber\\
&\null&\left.\left.\hskip-0.3in{\mathcal{H}}^2\right)(\psi+\Phi)+2{\mathcal{H}}\frac{d\Phi}{d\eta}+
4{\mathcal{H}}\frac{d\psi}{d\eta}\right]\delta_{\alpha\beta}+
\partial_\alpha \partial_\beta(\psi+\Phi)\;.\;
\end{eqnarray}
In these expressions, ${\mathcal{H}}$ is the Hubble parameter written as ${a^{'}}/{a}$.

		\subsection{The Perturbed Stress-Energy Tensor}
		The perturbed stress-energy tensor for a perfect fluid may be written 
		\begin{equation}
		{T^a}_b=(\rho+\delta\rho+p+\delta p)(u^a+\delta u^a)(u_b+\delta u_b)+ (p+\delta p)\delta^a_b\;,
		\end{equation}
		where $\rho$ and $\delta\rho$ are the total, and perturbed, energy density of the Universe contributed by all the species, 
		$p$ and $\delta p$ are the total, and perturbed, pressure contributed by all the species, and $u_a$ and $\delta u_a$ are
		the total, and perturbed, four-velocity. Then, using $g_{ab}\,u^au^b=-1$, the perturbed velocity components are 
		\begin{equation}
		\delta u^0=-\frac{\Phi}{a}\;,
		\end{equation}
		and
		\begin{equation}
		\delta u^\alpha=\frac{v^\alpha}{a}\;,
		\end{equation} 
		and we use the scalar-vector-decomposition to put $v_\alpha=\partial_\alpha v+v_\alpha^\perp$. 
		
		Working out the spatial components of Einstein's equation, i.e., $G_{\alpha\beta}=8\pi\,Gg_{\alpha c}T^c_\beta$, 
		we get
		\begin{eqnarray}
		\left[\nabla^2(\Phi-\psi)\right.&\hskip-0.3in+\hskip-0.3in&\left.2\frac{d^2\psi}{d\eta^2}+2\left(2\frac{d{\mathcal{H}}}{d\eta}+
		{\mathcal{H}}^2\right)(\psi+\Phi)
		+2{\mathcal{H}}\frac{d\Phi}{d\eta}+\right.\nonumber\\
		&\null&\left.\hskip-0.5in 4{\mathcal{H}}\frac{d\psi}{d\eta}\right]
		\delta_{\alpha\beta}+\partial_\alpha\partial_\beta(\psi+\Phi)=
		8\pi Ga^2\left[\delta p-2\psi p\right]\delta_{\alpha\beta}\;.
		\end{eqnarray}
		When $\alpha\neq\beta$, we get $\psi=-\Phi$, and using this condition for the rest of the components in
		Einstein's equation, we find that 
		\begin{equation}
		2\frac{d^2\Phi}{d\eta^2}+4\bigg(2\frac{d{\mathcal{H}}}{d\eta}+{\mathcal{H}}^2\bigg)\Phi+
		6{\mathcal{H}}\frac{d\Phi}{d\eta}= 8\pi Ga^2(\delta p-2p\Phi)\;,
		\end{equation}
		\begin{equation}
		2\nabla^2\Phi-6{\mathcal{H}}\frac{d\Phi}{d\eta}=8\pi Ga^2(\delta \rho +2\rho \Phi)\;,
		\end{equation}
		and
		\begin{equation}
		2\partial_\alpha\frac{d\Phi}{d\eta}+2{\mathcal{H}}\partial_\alpha\Phi=8\pi Ga^2(\rho+p)v_\alpha\;,
		\end{equation}
		where $v_\alpha=\partial_\alpha v+v_\alpha^\perp$.
		
		Finally, using a wave decomposition with a Fourier series, defined for an arbitrary function as 
		\begin{equation}
		F_k=\frac{1}{(2\pi)^3}\int F(x)\,e^{i\vec{k}\cdot\vec{x}}\,d^3x\;,
		\end{equation}
		we get 
		\begin{equation}
		k^2\Phi_k+3{\mathcal{H}}\bigg(\frac{d\Phi_k}{d\eta}+{\mathcal{H}}\Phi_k\bigg)=4\pi\, Ga^2\delta \rho_k\;.
		\end{equation}
		$\delta \rho_k$ in the above equation comprises of the components which deviate from ideal cosmic fluid. In the case of $\Lambda$CDM $\delta \rho_k$ comprises of dark matter, baryons, neutrinos and radiation, whereas in the case of $R_{\rm h}=ct$ Universe, this term comprises of dark matter, dark energy, baryons, neutrinos and radiation.

	\section{The Perturbed Boltzmann Equations in $\Lambda$CDM}
	Equation~(37) describes the evolution of the perturbed gravitational potential $\Phi$ with a source term $\delta \rho$ 
	driving its growth. To solve it, one must determine the evolution equation for $\delta \rho$ as well. In 
	principle $\delta \rho$ represents the total perturbed energy density of the universe, comprised of perturbations of 
	individual species that clump above the smooth background. In this section, we shall obtain the evolution equations 
	for individual species by perturbing their corresponding Boltzmann equations.
	
	The Boltzmann equation for a given species (i.e., dark matter, baryons, radiation, etc.) describes the evolution 
	of its distribution function in the 8-dimensional phase space with 4-dimensions representing the four space-time 
	coordinates, $x^{\,i}$, and the other four comprising the 4-momentum, $p^{\,i}$. Using the additional constraint 
	$g_{ij}p^{\,i} p^{\,j}=-m^2$, this 8-dimensional phase space is reduced to 7-dimensions. In this work, we 
	choose the four spacetime coordinates, $x^{\,i}$, the magnitude of the 3-momentum, $p\equiv |\vec{p}|$, and its 
	direction, $\hat{p}^{\,\mu}$, as the independent variables. From Liouville's theorem, one obtains the following equation 
	for the evolution of the distribution function $f_s(x^{\,i},p,\hat{p}^{\,\mu})$ (for species `s'):
	\begin{eqnarray}
	\frac{df_s}{d\lambda}&=&\frac{\partial f_s}{\partial x^0}\frac{\partial x^0}{\partial \lambda}+
	\frac{\partial f_s}{\partial x^{\,\mu}}\frac{\partial x^{\,\mu}}{\partial \lambda}+
	\frac{\partial f_s}{\partial p}\frac{\partial p}{\partial \lambda}+
	\frac{\partial f_s}{\partial \hat{p}^{\,\mu}}\frac{\partial \hat{p}^{\,\mu}}{\partial 
		\lambda}\nonumber\\
	&=&C[f_s]\;,
	\end{eqnarray}
	where $\lambda$ and $C[f_s]$ are the affine parameter and the source/collision term for this species, respectively. 
	We define $P^{\,i}={dx^{\,i}}/{d\lambda}$, so that ${dx^{\,0}}/{d\lambda}=P^{\,0}$. Thus, dividing the above 
	equation by $P^{\,0}$, and neglecting the fourth term that is of second order, one gets 
	\begin{equation}
	\frac{df_s}{d\eta}=\frac{\partial f_s}{\partial \eta}+\frac{\partial f_s}{\partial x^{\,\mu}}
	\frac{P^{\,\mu}}{P^{\,0}}+\frac{\partial f_s}{\partial p}\frac{\partial p}{\partial \eta}=
	\frac{C[f_s]}{P^{\,0}}\;,
	\end{equation}
	where $\eta$ is the conformal time, in terms of the expansion factor $a(t)$ and cosmic time $t$. 
	Using ${P^{\,\mu}}/{P^{\,0}}=({p}/{E})\hat{p}^{\,\mu}$ (where $E$ is the energy) and geodesic equation, we may write 
	the above equation as follows:
	\begin{equation}
	\frac{dp}{d\eta}=-{\mathcal{H}}p+E\hat{p}^{\,\mu}\partial_{\,\mu}\frac{h_{00}}{2}-\frac{p}{2}
	\frac{dh_{\,\mu \nu}}{d\eta}\hat{p}^{\,\mu}\hat{p}^{\,\nu}\;,
	\end{equation}
	where $h_{\mu\nu}$ are the perturbed metric coefficients. Then substituting Equation~(40) 
	into Equation~(39), we get 
	\begin{equation}
	\frac{df_s}{d\eta}+\frac{p\hat{p}^{\,\mu}}{E}\frac{\partial f_s}{\partial x^{\,\mu}}+p\bigg(-{\mathcal{H}}+
	\frac{E}{p}\hat{p}^{\,\mu}\partial_{\,\mu}\frac{h_{00}}{2}-
	\frac{1}{2}h_{\,\mu \nu}^{'}\hat{p}^\mu\hat{p}^\nu\bigg)
	\frac{\partial f_s}{\partial p}=\frac{a}{E}(1-\Phi)C[f_s]\;.
	\end{equation}
	In arriving at this equation, we have substituted $P_0=\frac{E}{a}(1+\Phi)$, in terms of the
	perturbed gravitational potential, $\Phi$. 
	
	Next, we partition the distribution function into its unperturbed, $\bar{f}_s$, and perturbed, 
	${\mathcal{F}}_s$, components:
	\begin{equation}
	f_s(\eta, x^i, p, \hat{p}^i)=\bar{f}_s(\eta, x^i, p, \hat{p}^i)+
	{\mathcal{F}}_s(\eta, x^i, p, \hat{p}^i)\;.
	\end{equation}
	Thus, multiplying Equation~(41) by the energy and integrating over momentum space, and collecting 
	the zeroth-order terms, one gets 
	\begin{equation}
	\int \frac{d^3p}{(2\pi)^3}E(p)\frac{d\bar{f}_s}{d\eta}-\int \frac{d^3p}{(2\pi)^3}
	{\mathcal{H}}pE(p)
	\frac{\partial \bar{f}_s}{\partial p}=\int \frac{d^3p}{(2\pi)^3}aC[f_s]\;.
	\end{equation}
	In the context of $\Lambda$CDM, the quantity $C[f_s]$ is zero because the particle number is 
	conserved during the phase of fluctuation growth. Therefore, integrating the second term on 
	the left-hand side by parts and neglecting the boundary term, we arrive at the expression 
	\begin{equation}
	\int \frac{d^3p}{(2\pi)^3}E(p)\frac{d\bar{f}_s}{d\eta}+3{\mathcal{H}}\int\frac{d^3p}{(2\pi)^3}
	\bigg(E+\frac{p^2}{3E}\bigg)\,\bar{f}_s=0\;.
	\end{equation}
	This equation may be reduced further by using the following definitions of density and pressure: 
	\begin{equation}
	\rho_s=\int \frac{d^3p}{(2\pi)^3}E(p)\bar{f}_s\;,
	\end{equation}
	and
	\begin{equation}
	{\mathcal{P}}_s=\int\frac{d^3p}{(2\pi)^3}\frac{p^2}{3E}\bar{f}_s\;.
	\end{equation}
	If we now apply Equations~(44-46) to the dark matter fluctuation, we arrive at the expression
	\begin{equation}
	\frac{d\rho_{\rm dm}}{d\eta}+3{\mathcal{H}}(\rho_{\rm dm}+{\mathcal{P}}_{\rm dm})= 0\;.
	\end{equation}
	
	We multiply Equation~(41) by $E(p)$ and again integrate over momentum space using the 
	definition of $\rho$ and $\mathcal{P}$, but now collecting the first order terms: 
	\begin{equation}
	\frac{d(\delta \rho_{\rm dm})}{d\eta}+(\rho_{\rm dm}+{\mathcal{P}}_{\rm dm})
	\partial_{\,\mu}v^{\,\mu}_{\rm dm}+3{\mathcal{H}}(\delta \rho_{\rm dm}+
	\delta {\mathcal{P}}_{\rm dm})
	+3(\rho_{\rm dm}+ {\mathcal{P}}_{\rm dm})\frac{d\Phi}{d\eta}=0\;.
	\end{equation}
	We define $\delta_{\rm dm}\equiv{\delta\rho_{\rm dm}}/{\rho_{\rm dm}}$ and set both 
	${\mathcal{P}}_{\rm dm}$ and $\delta{\mathcal{P}}_{\rm dm}$ equal to zero. Then, 
	\begin{equation}
	\frac{d\delta_{\rm dm}}{d\eta}=\frac{1}{\rho_{\rm dm}}\frac{d(\delta\rho_{\rm dm})}{d\eta}-
	\frac{\delta_{\rm dm}}{\rho_{\rm dm}}\frac{d\rho_{\rm dm}}{d\eta}\;.
	\end{equation}
	Substituting this expression into Equation~(48) and decomposing them into Fourier modes, we find that
	\begin{equation}
	\frac{d\delta_{{\rm dm},\,k}}{d\eta} =-ku_{dm,k} -3\frac{d\Phi_k}{d\eta}\;.
	\end{equation}
	In arriving at this equation, we have also used $\partial_{\,\mu}v^{\,\mu}_{k}=ku_{dm,k}$, where $u_{dm,k}$ 
	is the velocity perturbation of dark matter. 
	
	Now, taking the second moment of Equation~(41) by multiplying it with $p\hat{p}^{\,\mu}$ and 
	contracting with $i\hat{k}_{\,\mu}$, and then integrating it over momentum, we get
	\begin{equation}
	\frac{d(\rho_{\rm dm}u_{{\rm dm},\,k})}{d\eta}+4{\mathcal{H}}\rho_{\rm dm}u_{{\rm dm},\,k}+
	k\rho_{\rm dm}\Phi_k=0\;,
	\end{equation}
	where $u_{{\rm dm},\,k}$ is the $k^{\rm th}$ velocity perturbation of dark matter. Again 
	substituting for ${d\rho_{\rm dm}}/{d\eta}$ in this equation, we get
	\begin{equation}
	\frac{du_{{\rm dm},\,k}}{d\eta}=-\frac{1}{a}\frac{da}{d\eta}u_{{\rm dm},\,k}-k\Phi_k\;.
	\end{equation}
	
	\section{The Perturbed Boltzmann Equations in $R_{\rm h}=ct$}
	The early $R_{\rm h}=ct$ universe consisted of $\approx 80\%$ dark energy and $\approx 20\%$ 
	radiation, with a small contamination of matter \cite{MeliaFatuzo2016}. At late times, these 
	fractions change to $\approx 67\%$ dark energy and $33\%$ matter. A coupling between dark matter 
	and dark energy is therefore unavoidable in this model, since particle number is not conserved. 
	The Boltzmann equations in this model will therefore be considerably different from those in 
	$\Lambda$CDM. The pertinent equations were derived earlier in \cite{Yennapureddy2019}, 
	but we summarize the procedure here for consistency and convenience. 
	
	Starting with Equation~(41), an integration over momentum space yields
	\begin{equation}
	\int \frac{d^3p}{(2\pi)^3}E(p)\frac{d\bar{f}_s}{d\eta}+3{\mathcal{H}}\int\frac{d^3p}{(2\pi)^3}
	\bigg(E+\frac{p^2}{3E}\bigg)\bar{f}_s =\int \frac{d^3p}{(2\pi)^3}aC[f_s]\;,
	\end{equation}
	where the principal difference between this result and Equation~(44) is that $C[f_s]$ on the righthand 
	side is zero for $\Lambda$CDM, but not for $R_{\rm h}=ct$. The interaction term cannot be zero in the 
	latter due to the coupling between dark energy and dark matter. Then, using the definition of density 
	and pressure for dark matter (from Equations~45 and 46), one gets 
	\begin{equation}
	\frac{d\rho_{\rm dm}}{d\eta}+3{\mathcal{H}}\rho_{\rm dm}=
	\int \frac{d^3p}{(2\pi)^3}aC[f_{\rm dm}]\;.
	\end{equation}
	
	The aforementioned transition in densities, e.g., $\rho_{\rm de}\approx 0.8\rho_{\rm c}$ after the Big
	Bang to $\approx 2\rho_{\rm c}/3$ more recently, and analogously for radiation and matter, may be modeled 
	simply using a straightforward empirical expression, 
	\begin{equation}
	\rho_{\rm dm}=({\rho_{\rm c}}/{3a^2}) \exp\bigg[-\frac{a_*}{a}\frac{(1-a)}{(1-a_*)}\bigg]\;,
	\end{equation} 
	to reflect the fact that, in $R_{\rm h}=ct$, the Universe transitioned from a radiation-dark-energy dominated 
	early phase to a matter/dark-energy dominated one at late times. In the above expression, $\rho_{\rm c}$ is the 
	critical density today, and $a_*$ represents the scale factor at matter radiation equality. Using this expression, 
	one may easily evaluate the collision/source term, 
	\begin{equation}
	C[\bar{f}_{\rm dm}]=\frac{{\mathcal{H}}E}{a}\bar{f}_{\rm dm}+\frac{{\mathcal{H}}E}{a^2}
	\bar{f}_{\rm dm}\bigg(\frac{a_*}{1-a_*}\bigg)\;,
	\end{equation}
	which describes the interaction between dark energy and dark matter. A fraction of dark energy must be 
	transformed (possibly via decay) into dark matter. Thus, the corresponding collision term for dark energy is 
	\begin{equation}
	C[\bar{f}_{\rm de}]=-\frac{{\mathcal{H}}E}{a}\bar{f}_{\rm dm}-\frac{{\mathcal{H}}E}{a^2}
	\bar{f}_{\rm dm}\bigg(\frac{a_*}{1-a_*}\bigg)\;.
	\end{equation}
	
	Returning now to Equation~(41), with the use of Equation~(56), we find for dark matter that 
	\begin{eqnarray}
	\frac{df_{\rm dm}}{d\eta}&\hskip-0.3in+\hskip-0.3in&\frac{p\hat{p}^{\,\mu}}{E}\frac{\partial f_{\rm dm}}{\partial x^{\,\mu}}+
	p\left(-{\mathcal{H}}+\frac{E}{p}\hat{p}^{\,\mu}\partial_{\,\mu}\frac{h_{00}}{2}-
	\frac{1}{2}h_{\mu \nu}^{'}\hat{p}^{\,\mu}\hat{p}^{\,\nu}\right) \frac{\partial f_{\rm dm}}
	{\partial p}=\nonumber\\
	&\null&\left[{\mathcal{H}}f_{\rm dm}+
	\frac{{\mathcal{H}}}{a}f_{\rm dm} \left(\frac{a_*}{1-a_*}\right)\right](1-\Phi)\;. 
	\end{eqnarray}
	We take the first moment of this equation by multiplying it with $E(p)$ and integrating over momentum 
	space. Collecting the first order terms yields
	\begin{eqnarray}
	\frac{d(\delta \rho_{\rm dm})}{d\eta}&\hskip-0.3in+\hskip-0.3in&(\rho_{\rm dm}+{\mathcal{P}}_{\rm dm})
	\partial_{\,\mu}v^{\,\mu}_{\rm dm}+3{\mathcal{H}}(\delta \rho_{\rm dm}+\delta 
	{\mathcal{P}}_{\rm dm})+\nonumber\\
	&\null&\hskip-0.5in 3(\rho_{\rm dm}+{\mathcal{P}}_{\rm dm})\frac{d\Phi}{d\eta}= 
	\left[{\mathcal{H}}+
	\frac{{\mathcal{H}}}{a}\left(\frac{a_*}{1-a_*}\right)\right]
	(\delta \rho_{\rm dm}-\rho_{\rm dm}\Phi)\;.
	\end{eqnarray}
	
	We also have, from the definition of $\delta_{\rm dm}$ and ${\mathcal{P}}_{\rm dm}=\delta{\mathcal{P}}_{\rm dm}=0$ 
	for dark matter, that 
	\begin{equation}
	\frac{d\delta_{\rm dm}}{d\eta}=\frac{1}{\rho_{\rm dm}}\frac{d(\delta\rho_{\rm dm})}{d\eta}-
	\frac{\delta_{\rm dm}}{\rho_{\rm dm}}\frac{d\rho_{\rm dm}}{d\eta}\;.
	\end{equation}
	And substituting for $d(\delta\rho_{\rm dm})/d\eta$ in this equation, and isolating the Fourier mode $k$, we find that 
	\begin{equation}
	\frac{d\delta_{{\rm dm},\,k}}{d\eta} =-ku_{dm,k} -3\frac{d\Phi_k}{d\eta}-{\mathcal{H}}
	\bigg[1+\frac{a_*}{a(1-a_*)}\bigg]\Phi_k\;.
	\end{equation}
	Then, taking the second moment of Equation~(58) by multiplying it with $p\hat{p}^{\,\mu}$, contracting it with 
	$i\hat{k}_{\,\mu}$, and integrating over momentum space, we find 
	\begin{equation}
	\frac{d(\rho_{\rm dm}u_{{\rm dm},\,k})}{d\eta}+4{\mathcal{H}}\rho_{\rm dm}
	u_{{\rm dm},\,k}+k\Phi_k\rho_{\rm dm}
	={\mathcal{H}}\bigg[1+\frac{a_*}{a(1-a_*)}\bigg]\rho_{\rm dm}u_{{\rm dm},\,k}\;.
	\end{equation}
	Substituting for ${d\rho_{\rm dm}}/{d\eta}$ we thus get
	\begin{equation}
	\frac{du_{{\rm dm},\,k}}{d\eta}=-\frac{1}{a}\frac{da}{d\eta}u_{{\rm dm},\,k}-k\Phi_k\;.
	\end{equation}
	
	\begin{figure}
		\centering
		\includegraphics[scale=0.7]{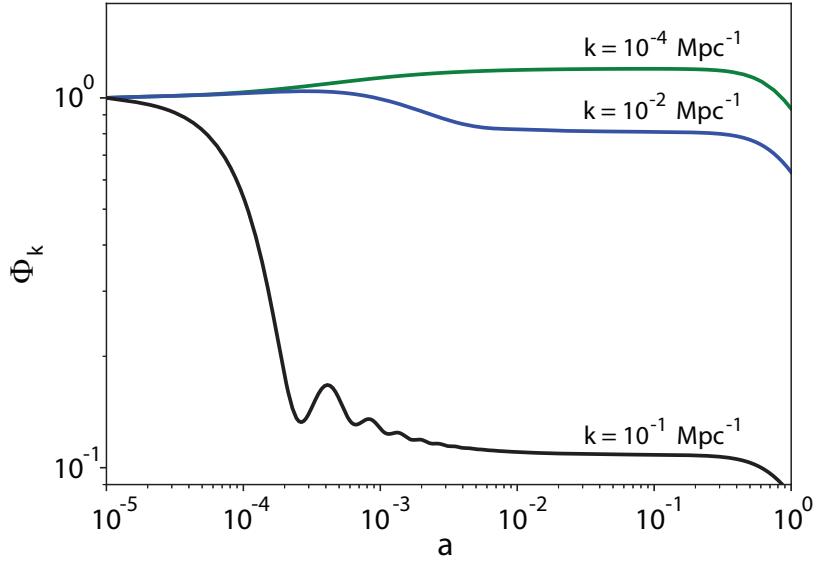}
		\vskip -0.2in
		\caption{Numerical solution of the perturbed potential, $\Phi$, in $\Lambda$CDM, for modes 
			$k=10^{-4}$, $10^{-2}$ and $10^{-1}$ Mpc$^{-1}$.} 
	\end{figure}
	
	A similar procedure allows to derive the analogous equation for dark energy. Using the interaction
	term in Equation~(57), we may write
	\begin{eqnarray}
	\frac{df_{\rm de}}{d\eta}&\hskip-0.3in+\hskip-0.3in&\frac{p\hat{p}^{\,\mu}}{E}\frac{\partial f_{\rm de}}{\partial x^{\,\mu}}+
	p\left(-{\mathcal{H}}+\frac{E}{p}\hat{p}^{\,\mu}\partial_{\,\mu}\frac{h_{00}}{2}-
	\frac{1}{2}h_{\,\mu \nu}^{'} \hat{p}^{\,\mu}\hat{p}^{\,\nu}\right) 
	\frac{\partial f_{\rm de}}{\partial p}=\nonumber\\
	&\null&-{\mathcal{H}}(1-\Phi)\left[1+
	\frac{a_*}{a(1-a_*)}\right]f_{\rm dm}\;,
	\end{eqnarray}
	where $f_{\rm de}$ is the distribution function for dark energy. Then, partitioning $f_{\rm de}$ 
	into its unperturbed ($\bar{f_{de}}$) and perturbed (${\mathcal{F}}_{de}$) components, we find
	\begin{eqnarray}
	\frac{d{\mathcal{F}}_{\rm de}}{d\eta}&\hskip-0.3in+\hskip-0.3in&\frac{p\hat{p}^{\,\mu}}{E}
	\frac{\partial {\mathcal{F}}_{\rm de}}{\partial x^{\,\mu}}-{\mathcal{H}}p\frac{\partial 
		{\mathcal{F}}_{\rm de}}{\partial p}+p\left(\frac{E}{p}\hat{p}^{\,\mu}\partial_{\,\mu}
	\frac{h_{00}}{2}-
	\frac{1}{2}h_{\,\mu \nu}^{'}\hat{p}^{\,\mu} \hat{p}^{\,\nu}\right)\frac{\partial \bar{f}_{\rm de}}
	{\partial p}=\nonumber\\
	&\null&-{\mathcal{H}}\left[1+
	\frac{a_*}{a(1-a_*)}\right]\left({\mathcal{F}}_{\rm dm}-
	\Phi\bar{f}_{\rm dm}\right)\;.
	\end{eqnarray}
	Taking the first momentum of this equation and isolating them into Fourier modes gives
	\begin{eqnarray}
	\frac{d \delta \rho_{\rm de}}{d\eta}&\hskip-0.4in+\hskip-0.4in&\rho_{\rm de}(1+w_{\rm de})\partial_{\mu}v^{\ \mu }_{de}+3{\mathcal{H}}
	\delta \rho_{\rm de}\left(1+\frac{\delta {\mathcal{P}}_{\rm de}}{\delta \rho_{\rm de}}\right)+\nonumber\\
	&\null&\hskip-0.5in3\frac{d\Phi}{d\eta}\rho_{\rm de}(1+w_{\rm de})=-{\mathcal{H}}\left[1+
	\frac{a_*}{a(1-a_*)}\right](-\rho_{\rm dm}\Phi+\delta \rho_{\rm dm})\;
	\end{eqnarray}
	From the definition of $\delta_{\rm de}={\delta \rho_{\rm de}}/{\rho_{\rm de}}$ and the inferred 
	equation-of-state ${\mathcal{P}}_{\rm de}=-\rho_{\rm de}/2$ \cite{MeliaAbdelqader2009} for dark energy, we may write
	\begin{equation}
	\frac{d\delta_{\rm de}}{d\eta}=\frac{1}{\rho_{\rm de}}\frac{d \delta \rho_{\rm de}}{d\eta}-
	\frac{\delta \rho_{\rm de}}{\rho_{\rm de}^2}\frac{d\rho_{\rm de}}{d\eta}
	\end{equation}
	so that, combining Equations~(66) and (67), we get 
	\begin{eqnarray}
	\frac{d\delta_{\rm de}}{d\eta}&\hskip-0.3in=\hskip-0.3in&-\frac{1}{2}\partial_{\mu}v^{\ \mu}_{de}-3{\mathcal{H}}\delta_{\rm de}
	\left(\frac{1}{2}+\frac{\delta {\mathcal{P}}_{\rm de}}{\delta \rho_{\rm de}}\right)- 
	\frac{3}{2}\frac{d\Phi}{d\eta}+\nonumber\\
	&\null&{\mathcal{H}}\left[1+\frac{a_*}{a(1-a_*)}\right]
	\frac{\rho_{\rm dm}}{\rho_{\rm de}}(\delta_{\rm de}-\delta_{\rm dm}+\Phi)\;.
	\end{eqnarray}
	
	We do not yet know the sound speed for the coupled dark matter/dark energy fluid, so we characterize
	it as follows:
	\begin{equation}
	c_s^2\equiv\frac{\delta {\mathcal{P}}}{\delta \rho}=\frac{\delta {\mathcal{P}}_{\rm de}}
	{\delta \rho_{\rm dm}+\delta \rho_{\rm de}}=\frac{\delta {\mathcal{P}}_{\rm de}/\delta 
		\rho_{\rm de}}{(1+\delta \rho_{\rm dm}/\delta \rho_{\rm de})}
	\end{equation}
	and, assuming adiabatic fluctuations, we shall put 
	\begin{equation}
	\frac{\delta {\mathcal{P}}_{\rm de}}{\delta \rho_{\rm de}}=c_s^2\bigg[1+\frac{2\rho_{\rm dm}}
	{\rho_{\rm de}}\bigg]\;.
	\end{equation}
	For the sake of simplicity, we assume the sound speed to be a constant limited to the range
	$0<(c_s/c)^2<1$. We have found that the actual value of this constant has a negligible impact
	on the solutions to the above equations since the ratio of dark matter density to dark energy
	is always much smaller than $1$ in the $R_{\rm h}=ct$ universe, and we therefore adopt the simple 
	fraction $c_s^2=c^2/2$ throughout this work. Thus, using Equations~(68) and (70), we get 
	\begin{eqnarray}
	\frac{d\delta_{\rm de,k}}{d\eta} &\hskip-0.3in=\hskip-0.3in& -\frac{k}{2}u_{{\rm de},\,k}-\delta_{\rm de,k}\left(
	\frac{3{\mathcal{H}}}{2}+3{\mathcal{H}}c_s^2+\frac{6{\mathcal{H}}c_s^2\rho_{\rm dm}}
	{\rho_{\rm de}}\right)-
	\frac{3}{2}\frac{d\Phi_k}{d\eta}+\nonumber\\
	&\null&{\mathcal{H}}\left[1+
	\frac{a_*}{a(1-a_*)}\right]\frac{\rho_{\rm dm}}{\rho_{\rm de}}(\delta_{\rm de,k}-
	\delta_{\rm dm,k}+\Phi_k)\;.
	\end{eqnarray}
	Finally, taking the second moment of Equation~(65), we find that 
	\begin{eqnarray}
	\frac{du_{{\rm de},\,k}}{d\eta}&\hskip-0.3in=\hskip-0.3in&-\frac{5{\mathcal{H}}}{2}u_{{\rm de},\,k}-k\Phi_k+2kc_s^2
	\left[1+\frac{2\rho_{\rm dm}}{\rho_{\rm de}}\right]\delta_{\rm de,k}+\nonumber\\
	&\null&{\mathcal{H}}\left[1+\frac{a_*}{a(1-a_*)}\right]\frac{\rho_{\rm dm}} 
	{\rho_{\rm de}}(u_{{\rm de},\,k}-2u_{{\rm dm},\,k})\;.
	\end{eqnarray}

\section{The Observed Matter Power Spectrum}
The matter power spectrum is obtained using several steps and a combination of data from the CMB, 
and the Ly-$\alpha$ forest, together with a model-dependent 
transfer function, that must be calculated individually for each cosmology. For the sake of clarity,
we begin by summarizing these key factors in the process.

\begin{figure}
	\centering
	\includegraphics[scale=0.7]{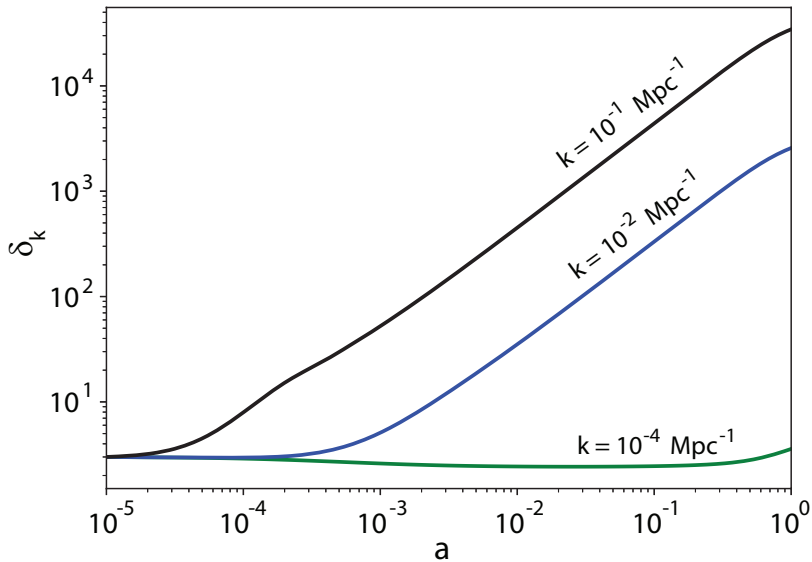}
	\vskip -0.2in
	\caption{Numerical solution of the matter fluctuation $\delta_{{\rm dm},\,k}$ in 
		$\Lambda$CDM, for modes $k=10^{-4}$, $10^{-2}$ and $10^{-1}$ Mpc$^{-1}$.} 
\end{figure}

\subsection{The Cosmic Microwave Background}
First and foremost, all of the data are model dependent, and must be recalibrated when changing
background cosmologies. For the concordance model parameter values, the CMB measurements are shown 
as blue, orange and black circular dots in Figure~3. The actual CMB observations are converted 
into a power spectrum using the approach of Tegmark \& Zaldarriaga (see ref. \cite{TegmarkZaldarriaga2002}). The CMB angular power 
for multipole $\ell$ may be written
\begin{equation}
C_\ell=\int_{-\infty}^{\infty}W_\ell(k)P_*(k)\,d\ln k\;,
\end{equation}
where $W_\ell(k)$ is the transfer function that depends on the cosmic matter budget and the 
reionisation optical depth, and $P_*(k)$ is the primordial power spectrum, assumed for simplicity, 
to be $P_*(k)\propto k$. These CMB measurements are mapped into the $k$-space of the matter power 
spectrum using the approach of Tegmark \& Zaldarriaga (see ref. \cite{TegmarkZaldarriaga2002}), the measured CMB data points $d_i$ (in the case of CMB they are $C_\ell$) are mapped into the probability distribution using the following equation 
\begin{equation}
d_i=\int_{-\infty}^{\infty}\mathcal{P}_i(k)d\ln k\;,
\end{equation} 

\begin{figure}
	\centering
	\includegraphics[scale=0.8]{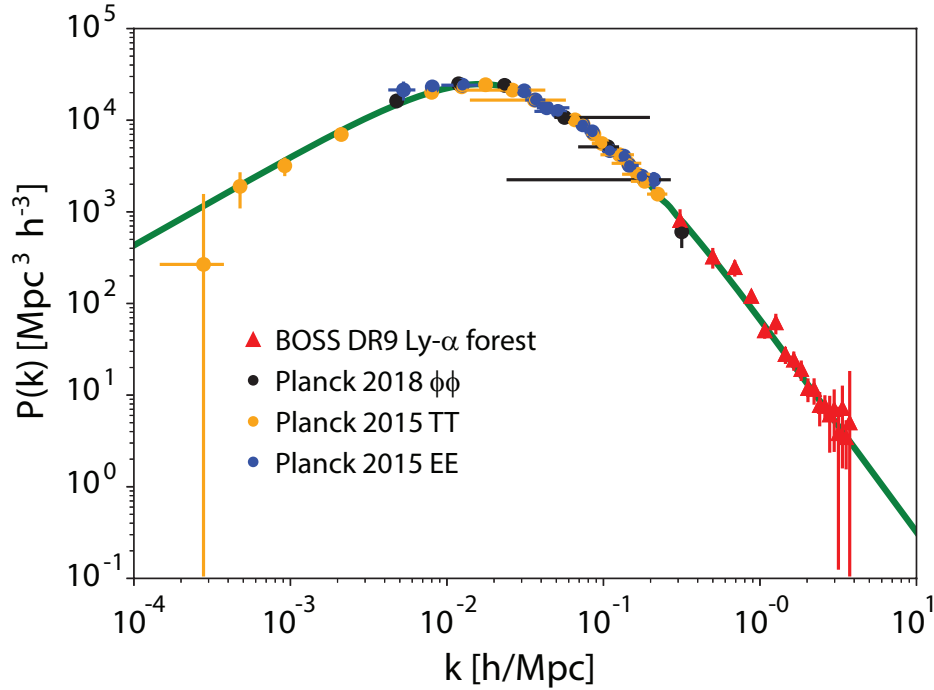}
	\caption{The matter power spectrum observed in the CMB (orange, blue and black dots), and the 
		Ly-$\alpha$ survey (red triangles), compared with the power spectrum predicted by 
		$\Lambda$CDM (solid black curve), where $h$ is the parameter $H_0/(100\;{\rm km}\;{\rm s}^{-1}
		\;{\rm Mpc}^{-1})$. Throughout this work, we assume the value $h=0.6732$, consistent with the
		latest {\it Planck} measurements (Planck Collaboration 2016).} 
\end{figure}
\noindent
where $\mathcal{P}_i$ is the probability distribution (see ref. \cite{TegmarkZaldarriaga2002}), in Figure~3, the CMB data points are placed horizontally at the $k$-value corresponding to the median of this distribution, with a horizontal error bar ranging from the 20th to the 
80th percentile of this distribution, representing a 1 $\sigma$ range. They are then plotted at a vertical position given by 
\begin{equation}
P_{\Lambda{\rm CDM}}(k)\simeq T^2_{\Lambda{\rm CDM}}(k_{\rm eff})P_*(k)\;,
\end{equation}
where $k_{\rm eff}$ is the horizontal location of the median of the window function and 
$T_{\Lambda{\rm CDM}}(k)$ is the matter transfer function in $\Lambda$CDM. In order to use these 
data for the $R_{\rm h}=ct$ universe, they must be recalibrated using the appropriate matter transfer 
function, according to the expression 
\begin{equation}
P_{R_{\rm h}=ct}(k)=P_{\Lambda{\rm CDM}}(k)\frac{T^2_{R_{\rm h}=ct}(k)}{T^2_{\Lambda{\rm CDM}}(k)}\;.
\end{equation}
The CMB data points recalibrated for $R_{\rm h}=ct$ are shown (also as blue, orange and black circular 
dots) in Figure~6.

\subsection{The Ly-$\alpha$ Forest}
The Ly-$\alpha$ forest (red triangles in Figs.~3 and 6) is produced by absorption seen in high 
redshift quasar spectra, associated with neutral hydrogen present in a continuously fluctuating 
photo-ionized intergalactic medium. Simulations show that the underlying mass density is related 
to the optical depth of the Ly-$\alpha$ absorption, which allows the Ly-$\alpha$ forest to be used 
as a proxy for the matter power spectrum. The gas traced by Ly-$\alpha$ is modestly overdense compared 
to the cosmic mean, so if all the relevant physics can be simulated, one may construct the underlying 
matter power spectrum from the observed spectrum. The TreeSPH hydrodynamical simulation that simulates 
the observed Ly-$\alpha$ forest indicates that the optical depth $\tau$ is proportional to the density 
of neutral hydrogen, according to the expression 
\begin{equation}
\tau(x)=A\rho_b(x)^\beta\;,
\end{equation} 
where $x$ is the line of sight distance towards the quasar and $A$ is the amplitude that depends on the 
cosmology and physical state of the gas. This amplitude $A$ is obtained by matching the simulated and
observed Ly-$\alpha$ forests. But given that the simulation depends on the assumed background cosmology, 
the value of $A$ is itself model dependent. There are several caveats in obtaining the matter power 
spectrum from these data, however. In particular, the hydro-simulations with a dark matter only 
prescription may not have included all of the relevant physics, and it is not clear how the uncertainties 
in the reionisation history, the ionizing background and its fluctuations propagate into the 
reconstruction of $P(k)$. These data may therefore not be as reliable as the others for model
selection purposes.

Nevertheless, we here attempt to overcome this model dependence as much as possible. The model dependence 
of the amplitude in the power spectrum arises from the process of matching the observed and simulated 
spectra. Thus, in order to properly compare the data with the predicted matter power spectrum in $R_{\rm h}=ct$,
one must carry out similar simulations with this model as the background cosmology. These simulations 
have not been performed yet, however, so we shall rely on a statistical argument to calibrate the matter 
power spectrum obtained from the Ly-$\alpha$ forest. The power spectrum obtained from the CMB 
and the Ly-$\alpha$ forest belong to the same sample, so one can use the $t$-test to examine whether the 
matter power spectrum obtained from the Ly-$\alpha$ forest through simulations is consistent with that 
obtained from the CMB. One may then also use the $t$-test to calibrate the matter power 
spectrum obtained from Ly-$\alpha$ by ensuring consistency with the spectrum obtained from CMB.

The matter power spectrum from the Ly-$\alpha$ forest may be calibrated using a two-step process. First, 
we optimize the polynomial function $f(k)= a_1k^{-1}+b_1k^{-2}+c_1k^{-3}+D_1$, where $a_1$, $b_1$, 
$c_1$ and $D_1$ are constants, by minimizing the $\chi^2$ in fitting the matter power spectrum from CMB. A second optimization procedure is followed using a polynomial $g(k)=a_2k^{-1}+b_2k^{-2}+c_2k^{-3}
+D_{2}$ to fit the matter power spectrum from Ly-$\alpha$, yielding the constants $a_2$, $b_2$, $c_2$ and 
$D_2$ via $\chi^2$ minimization. A relative calibration between these two is obtained by varying the 
normalization constant $D_2$ until the p-value obtained from the $t$-test lies above the $95\%$ confidence 
level. Using this approach, we find that the p-values for $\Lambda$CDM and $R_{\rm h}=ct$ are, respectively,
$\approx 99.8\%$ and $\approx 98.7\%$.

\section{The Matter Power Spectrum in $\Lambda$CDM}
In $\Lambda$CDM, dark energy is a cosmological constant, contributing to the smooth background 
but not the fluctuations. Thus, $\delta \rho$ in Equation~(37) is comprised primarily of matter, though radiation 
($\delta \rho_{\gamma}$) may contribute as well. But in this paper, we follow convention and consider only dark 
matter growth, so the radiation does not condense into the fluctuations. Like dark energy, it contributes only
to the smooth background. The equations describing the growth of dark matter fluctuations in $\Lambda$CDM may 
therefore be written (see Equations~37, 50 and 52) 
\begin{equation}
\frac{d\Phi_k}{d\eta}=-\bigg(1+\frac{k^2}{3{\mathcal{H}}^2}\bigg){\mathcal{H}}\Phi_k+
\frac{4\pi G a^2 \rho_{\rm m}}{3{\mathcal{H}}}\delta_{\rm m}\;,
\end{equation} 
\begin{equation}
\frac{d\delta_{{\rm dm},\,k}}{d\eta} =-ku_k -3\frac{d\Phi_k}{d\eta}\;,
\end{equation}
and
\begin{equation}
\frac{du_{{\rm dm},\,k}}{d\eta}=-\frac{1}{a}\frac{da}{d\eta}u_{{\rm dm},\,k}-k\Phi_k\;.
\end{equation}

Equation~(78) describes the evolution of the gravitational potential. The second term on the right-hand side 
in this expression represents the self-gravity driving the fluctuation growth, while the first is due to the 
so-called Hubble friction that suppresses growth. Equations~(79) and (80) describe the evolution of the matter 
fluctuations and their velocity. We solve these coupled first order differential equations starting with
$\delta_{m}=\frac{3}{2}\Phi$ and $u=\frac{1}{2}k\eta \Phi$, following the initial conditions used in 
the Cosmological Initial Conditions and Microwave Anisotropy Codes (COSMICS; see ref. \cite{Bertschinger1995}). As noted 
earlier, the expansion history in $\Lambda$CDM includes
an exponential acceleration that drives the modes across the Hubble horizon. They then freeze and later re-enter
as the Universe continues to expand. The small-scale modes re-enter during the radiation-dominated phase, while 
the large-scale modes re-enter when matter dominates. The modes re-enter across the Hubble horizon at different 
times, so the starting time when classical growth begins depends on their proper wavelength (or, equivalently, on
their comoving wavenumber $k$). Hence we choose the initial (conformal) time at the beginning of the fluctuation 
growth to be $\eta={\rm min}[10^{-3}k^{-1},10^{-1}h^{-1}Mpc]$ (COSMICS; see ref. \cite{Bertschinger1995}). The solution to these equations 
is plotted in Figures~1 and 2. Figure~1 shows the evolution of the gravitational potential $\Phi_k$, consistent
with (1) that all the modes are frozen outside the horizon, (2) that the small-scale modes that re-enter the 
horizon during the radiation-domination expansion decay at first, and grow only when matter starts to dominate, 
and (3) that the large-scale modes remain frozen outside the horizon during the radiation-dominated phase, and 
re-enter when matter dominates and grow continuously thereafter. 

Using the transfer function defined as $T(k,a)=\Phi_k(a)/\Phi_k(a_i)$ (see ref. \cite{Dodelson2003}), we then compute the matter power spectrum 
$P(k)=AP_*(k)T^2(k,a)$, in terms of the primordial power spectrum $P_*(k)$ generated by inflation, and the
normalization factor $A$. The computed matter power spectrum is shown in Figure~3, along with the power spectrum 
observed in the CMB, and Ly-$\alpha$. As noted earlier, the small-scale modes, 
$k\gtrsim 0.02$ Mpc$^{-1}$, re-enter the horizon during the radiation-dominated expansion and then decay,
as one may see to the right in Figure~3. The large-scale modes, $k\lesssim 0.02$ Mpc$^{-1}$, re-enter the 
horizon when matter is dominant and then grow. To appreciate the behavior to the right of the peak in this 
plot, consider two typical modes $k_1$ and $k_2$ with $k_1>k_2$. The first mode has a shorter wavelength and
therefore re-enters before the second one. Mode $k_1$ thus suffers a greater decay due to the large Hubble 
friction. Hence mode $k_1$ has less power than mode $k_2$. Starting from the small modes and progressing 
towards the larger ones, the power increases until $k\approx0.02$ Mpc$^{-1}$, which corresponds to the mode 
that re-enters the Hubble horizon exactly at matter-radiation equality. Modes with $k\lesssim0.02$ Mpc$^{-1}$ 
don't decay and instead grow during the matter-dominant era. In principle, these modes should all have the 
same power if the primordial power spectrum is completely scale invariant, but Figure~3 indicates that the
power declines as we progress towards even larger mode wavelengths, to the left of $k\approx0.02$ Mpc$^{-1}$. 
This happens because the primordial power spectrum $P_*=Ak^{n_s}$ is not exactly scale invariant; it has
a scalar spectral index $n_s=0.967$. Clearly, the turning point in the power spectrum is highly sensitive
to the energy content in the form of matter and radiation. Any adjustment to these quantities will alter
the location of the matter-radiation equality and therefore change the shape of the matter power spectrum.

\subsection{A Possible Failure of the Mechanism Generating the Matter Power Spectrum}
The exit and re-entry of modes across the Hubble horizon are paramount for generating the observed 
matter power spectrum in $\Lambda$CDM. This mechanism, however, relies on the existence of an
inflationary epoch. Should inflation eventually be disfavoured by the observations, $\Lambda$CDM 
would be unable to account for the observed matter power spectrum. It is worth mentioning in
this regard that the angular correlation function measured by the Cosmic Background Explorer 
(COBE)\cite{Wright1996,Hinshaw1996}, {\it Wilkinson} Microwave Anisotropy Probe 
(WMAP)\cite{Bennett2003,Spergel2003}) and particularly {\it Planck} (Planck Collaboration 2016)\cite{PlanckCollaboration2016}, have indicated a rather strong tension with 
the prediction of $\Lambda$CDM. Slow-roll inflation predicts significant large-angle correlations in 
the CMB fluctuations which, however, are not seen by these instruments. Large-angle correlations 
are disfavoured by the data at a confidence level exceeding $8\sigma$ \cite{MeliaCorredoira2018}. Cosmic variance could mitigate this problem only partially, so a resolution to this flaw
in the model is still elusive. Should this failure persist, it would herald, not only an internal 
inconsistency of $\Lambda$CDM, but also its inability to account for the observed matter power spectrum.  

\begin{figure}
	\centering
	\includegraphics[scale=0.7]{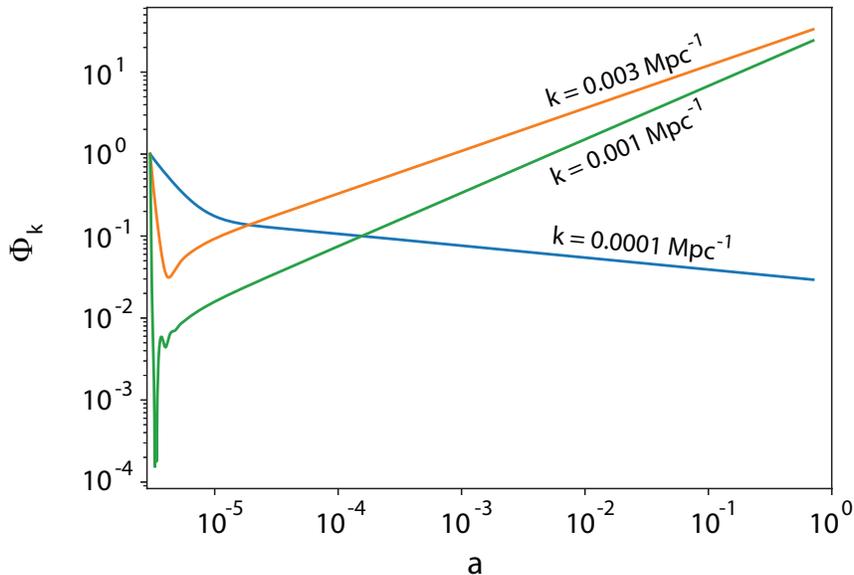}
	\vskip -0.2in
	\caption{Numerical solution of the perturbed potential $\Phi_k$ in $R_{\rm h}=ct$, for modes 
		$k=0.0001$, $0.003$ and $0.001$ Mpc$^{-1}$.} 
\end{figure}

\section{The Matter Power Spectrum in $R_{\rm h}=ct$}
The fluctuations in $R_{\rm h}=ct$ involve coupled species. The density $\delta\rho$ in Equation~(37) 
includes both dark energy and dark matter, i.e., $\delta\rho=\delta\rho_{\rm dm}+\delta\rho_{\rm de}$. 
Radiation and baryons are not directly coupled to dark matter, so we ignore their contribution in 
this paper. The relevant equations describing the growth of dark-matter fluctuations in 
$R_{\rm h}=ct$ may be derived as follows.

We convert the $\eta$ derivatives in the coupled first-order differential Equations~(37), 
(61), (63), (71) and (72) into derivatives with respect to the expansion factor $a$, and 
evolve  them starting from $a_i=10^{-12}$ to $a=1$. The evolution of $\Phi_k$ over this 
range is shown for several modes in Figure~4. This plot is to be compared with the corresponding
potential growth for $\Lambda$CDM in Figure~1.

The $\Lambda$CDM and $R_{\rm h}=ct$ models differ both in terms of how the 
perturbations are generated, and how they evolve. The perturbations in $\Lambda$CDM are 
seeded in the inflaton field, and then exit the horizon and freeze out during the
quasi-exponential expansion. These frozen modes later re-enter through the horizon as
the Hubble radius continues to grow, though at different times depending on the wavelength
of the fluctuation. In $R_{\rm hl}=ct$, on the other hand, the primordial fluctuations are 
generated within an incipient scalar field with an equation-of-state $p=-\rho/3$. Since
this model has no horizon problem \cite{Melia2013b}, it also does not experience any
inflated expansion, so none of the modes criss-cross the horizon as the Universe expands. 
Instead, all of the modes are born at the Planck scale, though at different times 
corresponding to their wavelength. 

The primordial power spectrum is produced from this birth mechanism, first proposed in 
a less well-defined manner by Hollands and Wald \cite{Holland2002,Melia2017}. The incipient 
scalar field should not be confused with the dark energy component, however, since the latter 
emerged after the former decayed into standard model particles and other fields seen in
extensions to the standard model of particle physics, e.g., representing dark energy. The 
incipient scalar field and dark energy have different equations-of-state. As described 
earlier, modes born during inflation freeze upon exiting the horizon and remain coherent 
upon re-entry, a  mechanism that is resposible for generating a near scale-invariant power 
spectrum. In the case of the incipient field in $R_{\rm h}=ct$, a similar outcome ensues 
upon exiting the Planck regime, but these fluctuations then remain within the semi-classical 
Universe and never criss-cross the horizon. They oscillate and grow until the Universe cools 
down to the GUT scale, and then presumably decay, as described above. 

The second difference between the two models arises during the evolution of the fluctuations, 
which may be understood via an inspection of Equation~(37). After introducing the model 
dependence through ${\mathcal{H}}$, $\rho_{\rm dm}$ and $\rho_{\rm de}$ in $R_{\rm h}=ct$, 
this expression may be written
\begin{equation}
\frac{d\Phi_k}{da}=-\bigg(1+\frac{k^2c^2}{3H_0^2}\bigg)\frac{\Phi_k}{a}+\frac{1}{2a}\bigg(\varpi(a) \delta_{\rm dm,k}+ \Upsilon(a)\delta_{\rm de,k}\bigg)\;,
\end{equation}
where $\varpi(a)=0.3\exp\left[-\frac{a_*}{a}\frac{(1-a)}{(1-a_*)}\right]$ and 
$\Upsilon(a)=1-\varpi(a)$. The corresponding expression in $\Lambda$CDM is 
\begin{equation}
\frac{d\Phi_k}{da}=-\bigg(1+\frac{k^2c^2}{3a^2H^2}\bigg)\frac{\Phi_k}{a}+\frac{3H_0^2}{2H^2a^2}
\frac{\Omega_{\rm m}}{a^3}\delta_{\rm dm}\;,
\end{equation}

\begin{figure}
	\centering
	\includegraphics[scale=0.7]{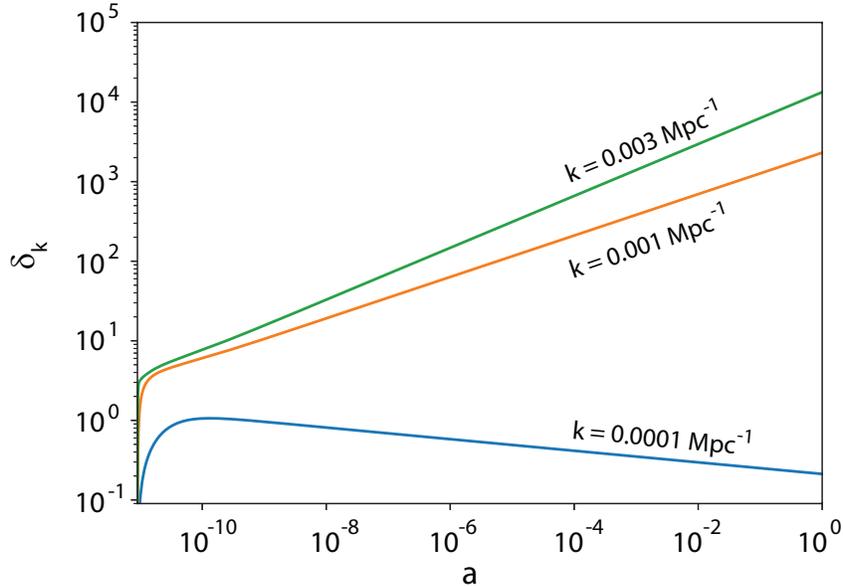}
	\vskip -0.2in
	\caption{Numerical solution of the perturbed $\delta_{dm,k}$ in $R_{\rm h}=ct$ for modes 
		$k=0.0001$, $0.003$ and $0.001$ Mpc$^{-1}$.} 
\end{figure}
\noindent
where $H=H_0\sqrt{\Omega_{\rm m}a^{\,-3}+\Omega_{\rm r}a^{\,-4}+\Omega_{\Lambda}}$. The Hubble
friction term ${k^2c^2}/{H_0^2}$ in $R_{\rm h}=ct$ may be written as ${4\pi^2R_{\rm h}^2}/
{\lambda^2}$, where $R_{\rm h}$ is the Hubble horizon and $\lambda$ is the proper wavelength 
of the mode. The primary difference between Equations~(81) and (82) enters via the ratio 
$R_{\rm h}/\lambda$. This factor is constant throughout the history of the Universe in 
$R_{\rm h}=ct$ for each given mode, while it evolves as a function of $a(t)$ in $\Lambda$CDM
for every mode. 

\begin{figure}
	\centering
	\includegraphics[scale=0.7]{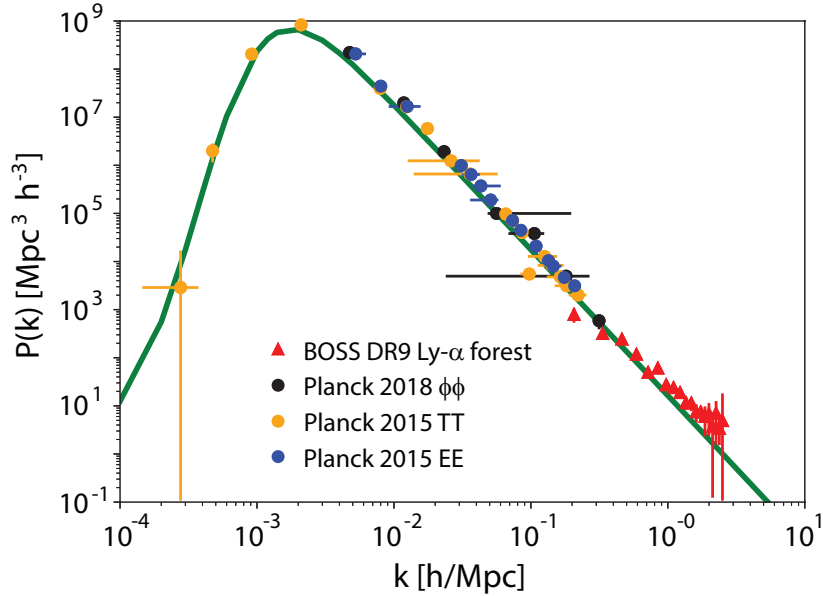}
	\caption{The matter power spectrum observed in the CMB (orange, blue and black dots), 
		and the Ly-$\alpha$ survey (red triangles), compared with the power spectrum predicted by $R_{\rm h}=ct$.} 
\end{figure}

The solution for $\delta_{dm,k}$ as a function of $a(t)$ in $R_{\rm h}=ct$ is shown in Figure~5, which is to be 
compared with Figure~2 for $\Lambda$CDM. The behavior of these solutions may be understood quite easily because 
the only `forces' acting on the modes are: (i) the inward pull of gravity, and (ii) the outwardly directed Hubble friction
due to the expansion of the Universe. As one can see in
Figure~4, the gravitational potential decays more rapidly for the smaller mode wavelengths,
i.e., the larger $k$'s. As a result, the growth rate of the modes varies according to their 
wavelength---the larger ones have a longer dynamical timescale, and therefore grow more slowly 
than the smaller ones. 

Another factor responsible for the difference between $R_{\rm h}=ct$ and the standard
model is that, at early times, the energy density in the former is a blend of dark energy and 
radiation, comprising $\sim 80\%$ and $20\%$, respectively. As the Universe expands, some of 
the dark energy decays into dark matter and standard model particles, eventually reaching a 
fractional representation of 2/3 dark energy and 1/3 matter (dark plus baryonic). The overall 
equation-of-state, however, always remains $p = -\rho/3$. Thus, both gravity and the pressure 
always act inwards together, making strucuture formation more efficient and more rapid than 
in $\Lambda$CDM. This is the principal reason why the growth rate in $R_{\rm h}=ct$ is larger
than in $\Lambda$CDM, and why it ends up matching the data much better than the latter.
For example, this strong growth rate is responsible for the early appearance of supermassive 
black holes and massive galaxies \cite{Yennapureddy2019}. The dynamical equations show that, 
as matter forms from dark-energy decay, the small-$k$ modes (i.e., the large-size fluctuations) 
overcome the Hubble friction due to the expansion and form bound structures. The large-$k$ 
modes (i.e., the small-size fluctuations), on the other hand, decay. This maybe seen more 
quantitatively in Fig.~5. Together with the transfer function $T(k,a)$ for $R_{\rm h}=ct$, 
we compute the power spectrum $P(k)$ using $P(k)=k^{n_s}T^2(k,a)$, where $n_s=0.967$. 
The combination of different growth rates in $\delta_{dm,k}$ and the $k$-dependent decay 
rate of the potential, produces the shape for the power spectrum seen in Figure~6. 

The computed power spectrum $P(k)$ in $R_{\rm h}=ct$ is shown in Figure~6, together
with the data recalibrated for this cosmological model (see \S~6 above). Above the peak,
the mode decay rate increases with $k$. As before, let us compare the behaviour of 
two particular modes, $k_1$ and $k_2$, with $k_1$ less than $0.001$ Mpc$^{-1}$ and the 
second mode $k_2$ greater than $k_1$ and $0.001$ Mpc$^{-1}$. The decay rate of $\Phi_{k_1}$ 
is smaller than that of $\Phi_{k_2}$ and the growth rate of $k_1$ is less than that of $k_2$, 
but mode $k_2$ did not have sufficient time to cross over $k_1$. This may be inferred by 
comparing Figures~4 and 5. Hence mode $k_1$ acquires more power than $k_2$. Now consider two 
modes $k_3$ and $k_4$, with $k_3$ less than $0.001$ Mpc$^{-1}$ and $k_4$ greater than $k_3$,
though still less than $0.001$ Mpc$^{-1}$. As $k_3$ is smaller than $k_4$, $\Phi_{k_3}$ 
undergoes less decay than $\Phi_{k_4}$, so the growth of $k_3$ is less than that of $k_4$, 
because the growth rate of $k_4$ exceeds that of $k_3$ and it had enough time to cross 
over. It ends up with more power than $k_3$. These effects all combine to produce the
particular shape of the power spectrum seen in Figure~6, particularly the height and location
of the peak. 

\begin{figure}
	\centering
	\includegraphics[scale=0.7]{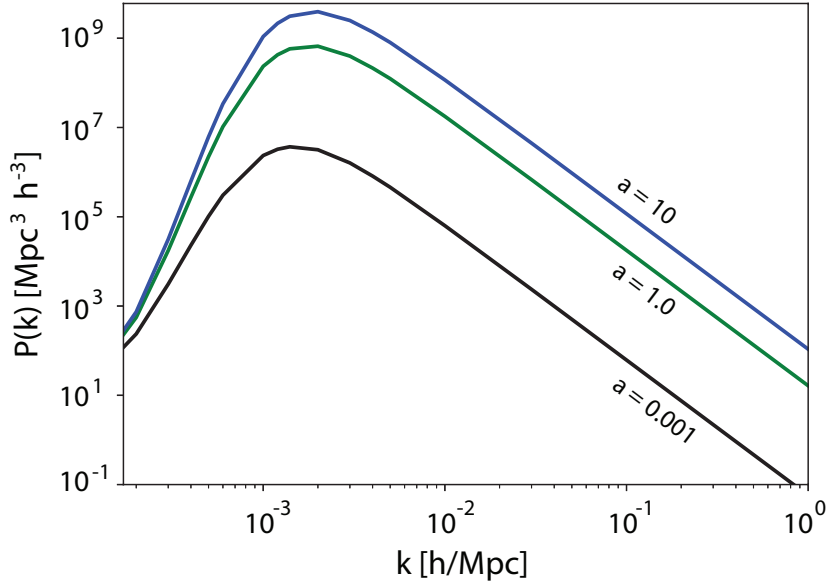}
	\caption{The matter power spectrum in $R_{\rm h}=ct$ calculated at different values
		of the expansion factor, specifically $a=10$, $1.0$ and $0.001$. Though the shape
		remains qualitatively the same, the peak of the distribution shifts slowly towards
		higher values of $k$ as the Universe ages.} 
\end{figure}

An inspection of Figure~7 demonstrates another important difference between the matter
power spectrum produced in $R_{\rm h}=ct$ versus its counterpart in $\Lambda$CDM. The
location of its peak is time dependent. The qualitative shape of this distribution
remains unchanged, but its turning point shifts as the Universe ages. The peak shifts 
slowly towards higher $k$ with time. Its predicted location at $a(t_0)=1$ (i.e., today)
matches the data very well. When viewed at $a=10$ (which in $R_{\rm h}=ct$ corresponds
to a Universe ten times older than today), however, the peak will have shifted from 
its current $k$ by a factor of almost $2$. In contrast, the peak's location does not
change as the $\Lambda$CDM Universe ages.

A second important factor to be noted in the case of $R_{\rm h}=ct$ is that the initial 
scale factor $a_i$ for evolving the modes is arbitrary in this approach; it does not pertain
to any particular epoch in the Universe's early expansion, such as the time at which the
modes re-enter the horizon, which does not happen in this model. In reality, the initial
value of $a(t)$ should correspond to the epoch at which the modes transitioned from the 
quantum realm to the classical realm. But the `classicalization' of these quantum fluctuations
is still far from being completely understood. We have chosen the value $a_i=10^{-12}$ 
to ensure that it does in fact lie beyond this transition. Of course, there is a wide range 
of possible values that could reflect this condition. To examine how sensitive our results are
to the choice of $a_i$, we have therefore repeated the calculation varying the starting
point, using $a_i=10^{-14}$ and $a_i=10^{-9}$. The results of this comparison are summarized
in Figure~8. Though the amplitude does change with $a_i$, the shape is only very minimally
dependent on it. But the amplitude is normalized to obtain the measured $\sigma_8$ today,
so this difference cannot be distinguished from one case to another. As one may see from
this figure, the predicted result is not at all dependent on the choice of $a_i$. The
principal physical reason behind this is that the transition from dark energy/radiation 
domination to dark energy/matter domination was optimized for $a_*=10^{-9}$ in this work. 
The growth of the modes is therefore dominated by what takes place after this point.
One may turn this approach around and view the `fitting' of the power spectrum as
a way of identifing the transition point $a_*$. 

\begin{figure}
	\centering
	\includegraphics[scale=0.6]{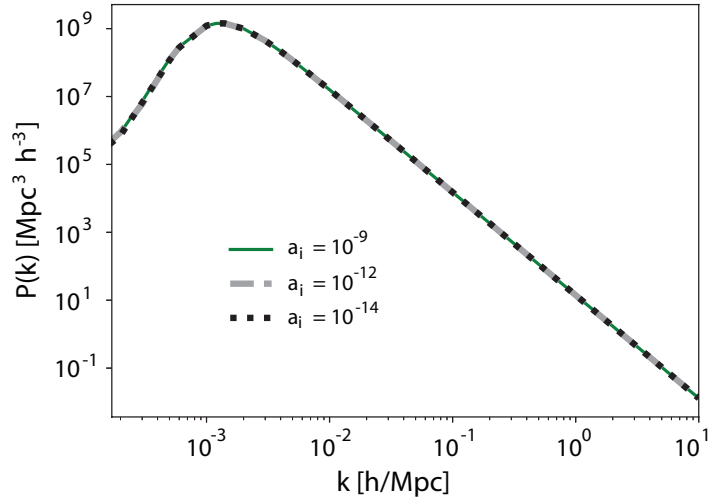}
	\caption{A comparision of the computed matter power spectrum for different values
		of the starting point $a_i$ in $R_{\rm h}=ct$.} 
\end{figure}

\section{Conclusion}
The mechanism responsible for the matter power spectrum in $R_{\rm h}=ct$ is quite
different from that developed for $\Lambda$CDM, primarily because the fluctuations
in this cosmology include a coupled dark matter--dark energy fluid. The formation of
this spectrum in $R_{\rm h}=ct$ is quite simple, largely influenced by the strong
dependence of the perturbed gravitational potential's ($\Phi_k$) decay rate on
the size of the modes: the smaller the scale size, the larger the decay rate. 
Coupled to the $k$ dependence of the mode growth rate, this trend shapes the
matter power spectrum, producing a distribution consistent with the observations.

It is important to stress that this mechanism does not require inflation at all, 
even to generate quantum fluctuations in the first place, since the presence of
any scalar field at the Planck scale would suffice \cite{Melia2017}. In contrast,
the formation of the matter power spectrum in $\Lambda$CDM requires the modes
to exit and re-enter the Hubble horizon at specific times in order to produce
the necessary growth and decay rates as functions of $k$. The physics of
inflation is still unknown, however, so this mechanism is arguably more
speculative than its counterpart in $R_{\rm h}=ct$. If the inflationary paradigm
turns out to be correct, then the principal effect responsible for shaping the
matter power spectrum is the $k$ dependence of the time at which modes re-enter
the horizon, with the smaller ones returning first, during the radiation-dominated
epoch, and the larger ones re-entering later, when matter was dominant. The
different growth rates during these epochs is the explanation for the peak and
shape of the distribution. The fact that such a complicated process is not needed
in $R_{\rm h}=ct$ is comforting to see, particularly since inflation is also not
needed to solve the temperature horizon problem, which does not exist in this model.

This distinction is important in view of the growing tension between $\Lambda$CDM
and the data in other areas. For example, with the discovery of the Higgs particle,
the standard model is now facing a second major horizon problem, this time having
to do with the electroweak phase transition \cite{Melia2018}. While the inflationary
(GUT) scale expansion may fix the temperature horizon problem, it cannot under
any known circumstance fix the subsequent electroweak horizon inconsistency. Coupled
to this, is the problem with the timeline in $\Lambda$CDM, which has significant
difficulty accounting for the `too early' appearance of massive halos and supermassive
black holes. With its much simpler structure formation process, the $R_{\rm h}=ct$ 
universe is able to solve ``the impossibly early galaxy problem" and the too early 
appearance of high-$z$ quasars. 

The fact that the cosmic fluid is dominated by dark energy throughout the Universe's
history in $R_{\rm h}=ct$, comprising $\approx 80\%$ of $\rho$ at early times and 
$2/3$ of $\rho$ towards the present, means that a coupling of dark matter and dark
energy is unavoidable in this cosmology. But though these fractions are constrained
quite tightly, the redshift at which this transition took place, gradual or otherwise,
is not known. In this paper, we have optimized the fit to the matter power spectrum
by treating the time at which this transition occurred as an adjustable parameter,
reflected in the value of $a_*$ in Equation~(55). Eventually, higher precision 
high-$z$ observations will be required to better constrain this process. Gamma-ray 
bursts (GRB) from Pop III stars \cite{BrommLoeb2007a,BrommLoeb2006a,BrommLoeb2006b} may provide 
such a diagnostic. In particular, the rate of gamma-ray 
bursts to be observed is strongly dependent on the growth rate, which itself
depends on the transition redshift (from dark energy/radiation to dark energy/matter
dominated epochs). The observation of GRB's by the James Webb Space Telescope 
(JWST) may prove to be critical for a proper comparison and evaluation of these 
two models.

\section*{Acknowledgments} 
We are grateful to Andreu Font-Ribera, David Weinberg and Arthur Kosowsky for very informative 
discussions regarding the observed matter power spectrum from Ly-$\alpha$ and the CMB. We are 
also thankful to Sol\`{e}ene Chabanier for sharing the matter power spectrum data with us. 
FM is grateful to the Institutode Astrofisica de Canarias in Tenerife and to Purple Mountain 
Observatory in Nanjing, China for their hospitality while part of this research was carried out.

\end{document}